# THE NEW GAS STRIPPER AND CHARGE STATE SEPARATOR OF THE GSI HIGH CURRENT INJECTOR

W. Barth, P. Forck GSI, Planckstr.1, 64291 Darmstadt, Germany

*Abstract*

The GSI Unilac was upgraded as a high current injector for the SIS in 1999. Therefore, the stripping section [1] at 1.4 MeV/u, where a beam transport under highest space charge conditions and multi beam operation had to be established, was completely new designed. Results of the commissioning of the stripper section will be presented. The beam transport to the new gas stripper, and the charge state analysis under space charge conditions confirmed the calculations. A $^{238}U^{4+}$ beam - coming from a MEVVA ion source and accelerated by the new injector linac- was stripped (with the expected stripping efficiency) to the charge state 28+ and successfully separated by the new spectrometer (15 degree and 30 degree kicker magnets). The transport and matching to the poststripper accelerator under highest space charge conditions was investigated with a 15 emA $^{40}Ar^{10+}$ beam. Especially space charge and charge state dependent emittance growth effects in 6d-phase space will be discussed.

## 1 INTRODUCTION

For the upgraded Unilac a new stripper section was designed and installed in 1999. Some additional design features had to be considered: charge state separation and beam transport under highest space charge conditions and multi beam operation with pulsed magnets.

## 2 SETUP OF THE STRIPPER SECTION

The layout is shown in Fig. 1. By two quadrupole doublets the beam is matched to the gas stripper. The new stripper device consists of the interaction zone (super-sonic $N_2$-jet) and three steps of differential pumping, upstream and downstream respectively. Compared to the old stripper [2], the free aperture in the new one is approx. 40% larger, ensuring a small beam size at the analyzing slit without any additional focusing elements. The charge state separator consists of three bending magnets, operating in pulsed mode. With the new $15^0$ fast kicker magnet (inflecting the HLI-beam from the ECR-source to the Unilac-axis) a multi-pulse mode from the different injectors is possible. In the following transport line the longitudinal matching (with two rebunchers at 36 MHz and 108 MHz) and the transversal matching (with a quadrupole doublet and a triplet) to the poststripper accelerator is accomplished. All sensitive elements are protected by diaphragms to handle with the highest beam pulse power along the Unilac (up to 1.4 MW).

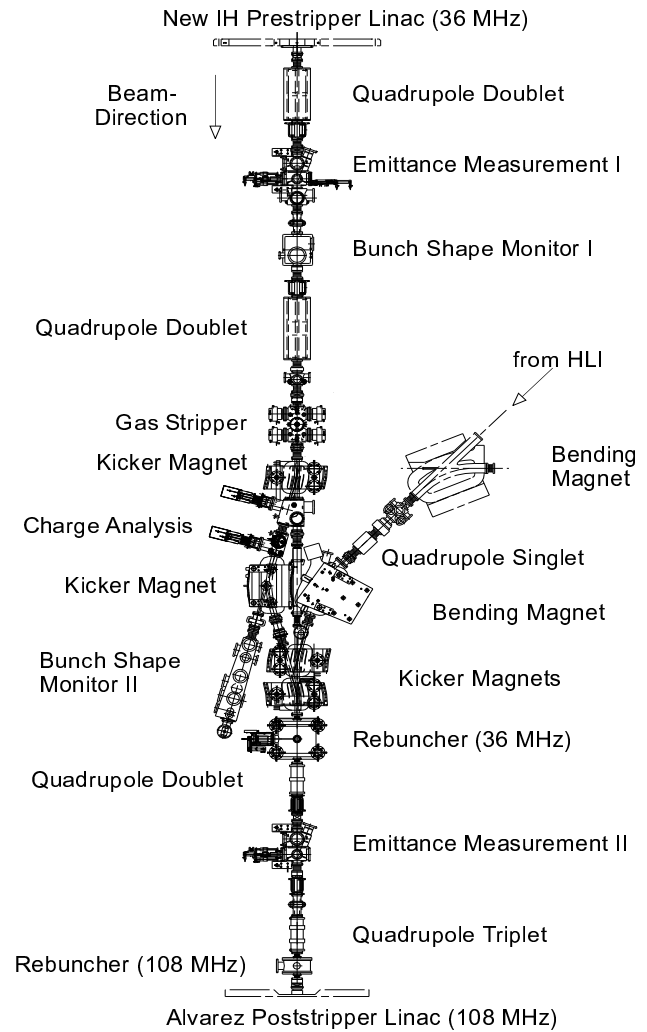

Fig. 1: Layout of the new 1.4 MeV/u stripper section (beam diagnostics is not completely represented); 14.0 m total length.

## 3 BEAM DYNAMIC

### 3.1 Matching to the gas stripper

The transverse emittance growth (compared to the "zero current" -value) after the prestripper accelerator for a space charge loaded $^{40}Ar^{1+}$-beam (up to 8 emA) generated in the MUCIS [3] ion source and accelerated in the HSI-LINAC is measured with the pepperpot measurement device. Beam size and divergence are in good agreement with calculations done with LORAS [4]. To match this beam to the gas stripper the beam is horizontally focused through the gas stripper with a convergence angle of 12 mrad resulting in a beam size for low intensity less than 5 mm for each charge state in the analyzing plane of the spectrometer.

## 3.2 Charge Separation

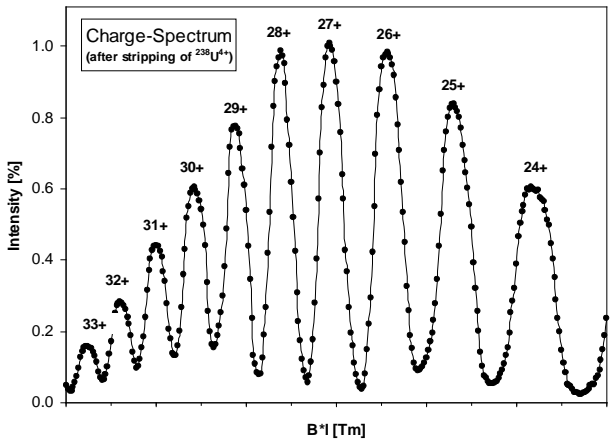

Fig. 2. Charge spectrum after stripping of a $^{238}U^{4+}$ beam (150 eµA)

After the replacement of the former gasjet stripper, an increase of the inlet pressure (about 50%) was necessary to provide the same stripping efficiency as before. A measured charge state spectrum after stripping of a low intensity $^{238}U^{4+}$ beam as shown in Fig. 2 yield in a stripping efficiency of 14 % for the main charge state (28+) as hitherto.

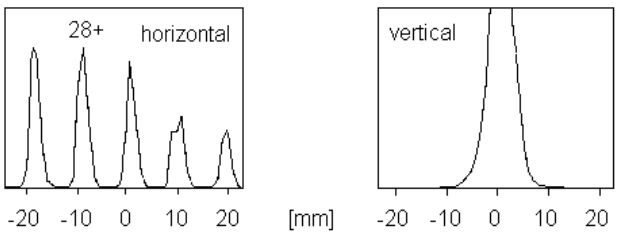

Fig. 3: Transverse profiles after stripping a $^{238}U^{4+}$ beam (I=150 µA).

Fig. 3 epitomizes the measured transversal beam profiles for a $^{238}U^{4+}$ beam (150 eµA) after stripping in the $N_2$-gasjet and after transport to the analyzing slit. The dispersion in the spectrometer is high enough to separate all charge states in the desired range of ion species without any beam loss in the main charge state. The vertical beam size is less than 20 mm, small enough to pass the spectrometer without particle loss.

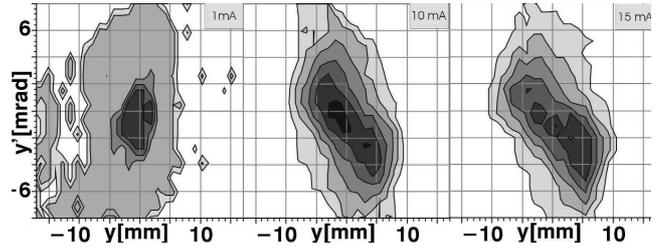

Fig. 4. Measured beam emittance as a function of intensity.

## 3.3 Matching to the poststripper

For the $Ar^{10+}$ beam Fig. 4 represents the variation of the beam emittance influenced by space charge forces due to an increase of the beam intensity. The intensity variation is done in the LEBT section. Whereas the beam divergence measured after the prestripper does not change, a significant dependence of the beam distribution after stripping and transport to the poststripper takes place. Thus intensity depending matching to the Alvarez accelerator is inevitable, if a high transmission rate for the whole Unilac is desired.

## 4 SPACE CHARGE EFFECTS

For space charge dominated beams no considerable change of the charge state distribution takes place. As shown in Fig.5 a max. $Ar^{10+}$ current of 18 emA is

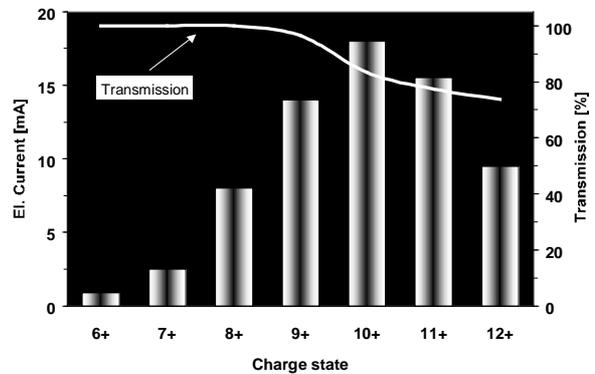

Fig. 5: Charge spectrum after stripping of an $^{40}Ar^{1+}$ beam (7.5 emA) under space charge conditions – the transmission represents the fraction of particles passing the spectrometer.

received, after the stripping of an $^{40}Ar^{1+}$ beam - thus the relative fraction is approx. 25 %. The transport at design intensity of 6 emA is particle loss free. The threefold intensity (18 mA) was reached after stripping with some decrease of the particle transmission (85 %) in the spectrometer, in contrast to the more pessimistic beam dynamic simulations. This is possibly a hint to a higher beam neutralization rate after stripping, resulting in lower emittance growth and higher transmission. For higher charge states the particles are more affected by space charge forces, especially in the drift section to the

spectrometer, where the electric beam current is the highest in the whole Unilac. The particle transmission for higher charge states decreases as shown in Fig. 5.

### 5.1 Transversal emittance growth

For measurements of the absolute horizontal emittance the already existing analyzing slits in the region of highest dispersion (after the first dipole of the charge state separator) were used, together with a profile grid two meters behind to complete for another emittance measurement device. A measured 90% emittance of 0.49 π·mm·mrad for the "zero current case" is in good agreement with calculation done with LORAS and PARMILA-Transport for 5 mA ($Ar^{1+}$) and without any space charge effects after stripping. As shown in Fig. 6 the emittance increases significantly as a function of the $Ar^{10+}$-intensity, but less than expected by calculation. At the design intensity of 6 mA an emittance growth of less than 10% was observed, in contrast to 40% as expected.

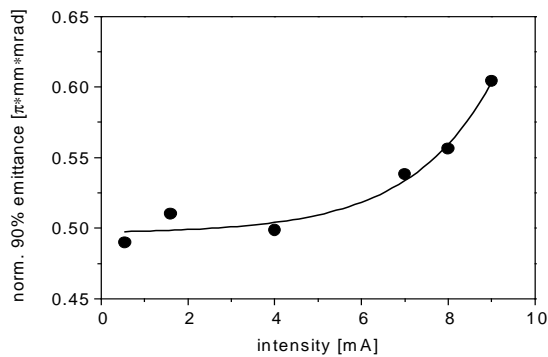

Fig. 6: Horizontal emittance as a function of the $Ar^{10+}$ beam intensity; attenuation is done by stripper density variation (5mA $Ar^{1+}$ input intensity).

### 5.2 Bunch shape measurements

Results of the measurements done with the bunch shape monitors II [5] with a $U^{4+}$ ion beam are resumed in Fig. 7. Without any space charge effects in the HSI a typical bunch shape stemming from the RFQ is occurred. Admitting a higher beam current in the HSI results in smaller bunch shape. If stripping and transport under space charge conditions is done the bunch width increases by an factor of 2.

## 5 SUMMARY

The redesigned stripper section of the new High Current Injector was installed and successfully commissioned in 1999. The attached beam diagnostic devices are suited to investigate the beam properties in the section. Matching to the stripper, stripping by itself, charge state separation and transport to the Alvarez accelerator were verified and in good agreement with multi-particle calculation over a wide range of beam intensities. The beam quality is as good as expected. For the stripping of $U^{4+}$ as a worst case the charge state separation is excellent. The measurements of the transverse beam profile, emittance and bunch shape indicate a partly space charge compensated beam after the stripping process.

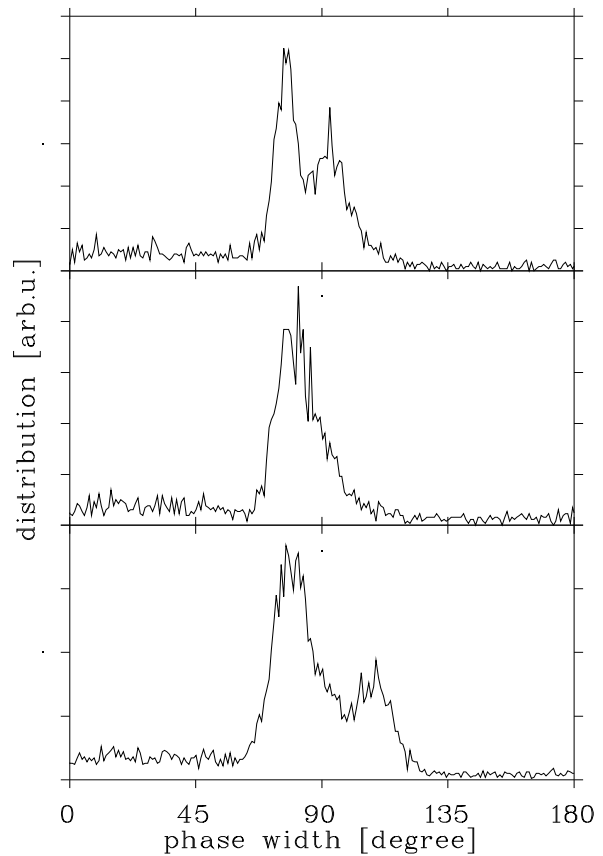

Fig. 7: Bunch shape measurements with an $U^{28+}$ beam after stripping: low intense HSI beam (above), high intense HSI beam and low after stripping (middle), high intensity in the HSI and after stripping (beneath).

## REFERENCES

[1] W. Barth, J. Glatz, J. Klabunde, U. Ratzinger, "Space Charge Dominated Beam Transport in the 1.4 MeV/u-Unilac stripper section", LINAC'96, Geneva, August 1996.
[2] J. Klabunde, W. Barth, L. Dahl, J. Glatz, Unilac Status and Developments, GSI Scientific rep. 1997, p. 159
[3] R. Keller, Multicharged ion production with MUCIS, GSI Scientific rep. 1987, p. 385
[4] U. Ratzinger, The new GSI Prestripper Linac for high current heavy ion beams, LINAC96, Geneva, Switzerland, p. 288 (1996)
[5] P. Forck, F. Heymach, U. Meyer, P. Moritz, P. Strehl, Aspects of Bunch Shape Measurements for Slow, Intense Ion Beams, DIPAC 1999, Chester